\documentclass[12pt,aps]{revtex4}
\begin{document}
\title
{Exactly solvable spin dynamics of an electron coupled to large number of
nuclei and the electron-nuclear spin echo in a quantum dot}
\author{Kozlov G.G.}

\begin{abstract}

The model considered in the paper is used  nowadays to describe spin dynamics
of quantum dots after optical excitation.
Based on the exact diagonalization of a model Hamiltonian, we solve the problems
 of the electron spin polarization decay and magnetic field dependence of the
 steady state polarization.
 The important role of the nuclear state is shown and methods of its
 calculation  for different regimes of optical excitation are proposed.
 The effect of spin echo observed after application of the magnetic field
 $\pi$-pulse is predicted.
\end{abstract}
\maketitle 

\hskip100pt {\it e}-mail:  gkozlov@photonics.phys.spbu.ru

\section{Introduction}

The system comprised of an electron spin coupled to a large number of nuclear
spins is a popular object  of radio spectroscopy.
The relevant mathematical model related to high magnetic fields typical for the
 EPR and NMR spectroscopy is studied fairly well.
   The recent interest to this model with weak magnetic fields is associated
   with appearance of new physical objects (quantum dots)
 whose spin dynamics may be described, with a sufficient accuracy, by this
 model.
The up-to-date laser technic makes it possible to prepare the
electron spin in a quantum dot in the state with a specified
projection, while observation of the polarized luminescence allows
one to detect the electron spin dynamics. General picture of the
electron spin dynamics in these experiments is described in
sufficient detail in the appropriate theoretical papers
 (\cite{Sch,Sem,Coi,Cha,Cha1,Sig,Kha,Kha1,Ste}). The spin Hamiltonian used in these papers is
well known in spectroscopy of free radicals and similar systems.
 At the moment of our first publication of this paper in arXiv, we were not
 aware of any papers that describe and use the fact that, in the case of
 uniform electron density on nuclei, such a Hamiltonian can be diagonalized
 exactly.
 This is why, in the first version of this publication, we presented solution
 of this problem.
  However, just a few days after our publication, we received a reference to the
 paper \cite{Zha} in which this result was described.
  Thus, the results presented below in section 2,  in fact, have been
 already obtained earlier in \cite{Zha}.
 Note also that the
system treated below is characterized by a large number of the
degrees of freedom, and exact solution of the relevant kinetic
problems may be of independent interest.

The system under consideration is comprised of electron spin $1/2$
 coupled to even number $2N$ of nuclei with spins $1/2$.
Interaction of the electron is supposed to be the same for all the nuclei.
 The Hamiltonian of the problem has the form:
   \begin{equation}
   H=\omega S_z +A_{\|}S_z I_z+A_{\bot}(S_+I_-+S_-I_+)
   \end{equation}
   \begin{equation}
   {\bf I}=\sum_{\alpha=1}^{2N} {\bf I}_\alpha,\hskip10mm I_\alpha^2={3\over 4}
   \end{equation}

Hear $S_z$ и $I_z$ --are the $z$-projections of the electron spin and total
nuclear angular momentum,
$S(I)_\pm=S(I)_x\pm \imath S(I)_y$ --
 are the standard lowering and raising operators for the $z$- projection of the electron ( total
 nuclear) spin,  $\omega$ --
 is the external magnetic field expressed in frequency units, $A_\|$ и $A_\bot.$ --
 are the components of the hyperfine coupling tensor.
 The main results of the paper are:

 1. It is shown that Hamiltonian (1) can be diagonalized exactly.

 2. The formulas that describe dynamics of the initially polarized electron spin
 for a given initial nuclear density matrix are derived.
 Note again that this result was presented in\cite{Zha}.

 3. Formulas that describe pulse-to-pulse transformation of the nuclear density
matrix for periodic orientation of electron spin by optical pulses, which is  typical for experiments with
quantum dots, are obtained.

  4. The calculated steady-state nuclear density matrices for the regimes of
constant-sign and sign-alternating optical orientation of the electron spin.
It is shown that, in the case of constant-sign orientation, a nuclear
polarization arises whose ultimate value is calculated.

5. It is shown that typical experimental regimes of the electron spin  optical
orientation may be accompanied by a certain ordering of the nuclear density
matrix with respect to the total nuclear angular momentum and its projection.
The latter is reveal in quasi-periodicity of the electron spin dynamics,
 which is absent in the case of the high-temperature nuclear density matrix.

 6. The examples of calculations  of the electron spin dynamics and magnetic-field
dependence of its steady-state polarization for various regimes of optical
orientation are presented.

 7. The echo effect in the dynamics of the electron spin in a quantum dot after a $\pi$-pulse
 of the magnetic field is predicted.

 \section{Diagonalization of the Hamiltonian}

Hamiltonian (1) depends on the total angular momentum of all the nuclei.
So, it makes sense to employ the representation of wave functions that are
eigen functions the operators of total nuclear angular momentum squared and its
 $z$-projection.

These functions have the following quantum numbers:

 $I$ --
 is the total angular momentum quantum number related to the mean value of
 the angular momentum squared by the formula:
  $\langle I^2\rangle=I(I+1)$ (in the units of $\hbar^2$),

$L$ -- $z$-projection of the total nuclear angular momentum  (in the units of $\hbar$).

This set is, however, not complete because one and the same value of the total
nuclear angular momentum can be obtained by summing up  the elementary
nuclear angular momenta in many different ways.
 Let us number the set of the remaining quantum numbers by the index $\alpha$:
$\Psi_{I,L,\alpha}\equiv |I,L,\alpha\rangle$.
 In this representation the matrices of the total nuclear angular momentum
squared and its $z$-projection are diagonal.
 Note that for the states with a given quantum number of the total nuclear angular
 momentum squared $I$, the diagonal elements of its matrix are $I(I+1)$
  while the diagonal elements of $I_z$ are $L=-I,1-I,...,I-1,I$.
Thus the dimesionality of the subspace with given number $I$ and $\alpha$
equals $2I+1$.
As is known from the theory of angular momentum, the matrices of the $x$- and
$y$- projections of the total angular momentum $I_x$  and $I_y$ (and, therefore, $I_\pm$ )
have no nonzero matrix elements between the states with non equal quantum
number of the total angular momentum: $\langle I,...|I_{x,y}|I',...\rangle=0$
 for $I\ne I'$.
 It follows from the fact that the operator of the total angular momentum
squared commutes with the operator of any of its projection.
 Let us show that the set  of states $|I,L,\alpha\rangle$  can be chosen in such
a way that the operators $I_\pm$ will have zero matrix elements also between
the states differing only by the quantum number $\alpha$ even for the same $I$.
 Consider, for  this purpose the states with a given total angular momentum $I$
and the maximum $z$-projection $L=I$ at different $\alpha$: $|I,I,\alpha\rangle$.
 The number of these states will be equal to the number of ways that allow one
to obtain the total angular momentum $I$ by summing up $2N$ angular momenta
 1/2.
 These states are linearly independent and, without the loss of generality, they
can be regarded as normalized and mutually orthogonal:
$\langle I,I,\alpha|I,I,\alpha'\rangle=\delta_{\alpha\alpha'}$..
 We obtain all the rest states  from these ones using subsequently the
lowering operator for $z$-projection:
\begin{equation}
|I,L,\alpha\rangle=I^n_-|I,I,\alpha\rangle,\hskip10mm L=I-n
\hskip10mm n=0,1,...,2I
\end{equation}
 Let us show that the set of functions obtained in this way is orthogonal.
 Each of these functions, in conformity with the theory of angular momentum,
  is the eigen function of the operator of the total angular momentum
  $z$-projection with the eigen value $L$.
 For this reason, the functions with different quantum numbers of the total
angular momentum $z$-projection will be necessarily orthogonal as eigen functions of a
self-conjugate operator corresponding to different eigen numbers.
  What is left is to show orthogonality of the functions differing only by the
 quantum number $\alpha$. So, we have:

\begin{equation}
\langle I,I-n,\alpha| I,I-n,\alpha'\rangle=
\langle I,I,\alpha|I_+^nI_-^n |I,I,\alpha'\rangle=
\langle I,I,\alpha|I_+^{n-1}I_+I_-I_-^{n-1}|I,I,\alpha'\rangle
\end{equation}

In conformity with the theory of angular momentum, the function $I_-^{n-1}|I,I,\alpha'\rangle$
 will be necessarily eigen function for the operators of the total angular
 momentum squared and its projection, with the eigen numbers being equal,
 respectively, to $I(I+1)$ и $L=I-n+1$.

As a result, taking into account the identity

\begin{equation}
I_+I_-={\bf I}^2-I_z^2+I_z,
\end{equation}
we have
\begin{equation}
I_+I_-I_-^{n-1}|I,I,\alpha'\rangle=
\bigg[I(I+1)-L^2+L\bigg] I_-^{n-1}|I,I,\alpha'\rangle\bigg|_{L=I-n+1},
\end{equation}
 and, therefore,

\begin{equation}
\langle I,I-n,\alpha| I,I-n,\alpha'\rangle\sim
\langle I,I,\alpha|I_+^{n-1}I_-^{n-1}|I,I,\alpha'\rangle
\end{equation}

Repeating this reasoning $n-1$ times more, we obtain:

\begin{equation}
\langle I,I-n,\alpha|I,I-n,\alpha'\rangle\sim
\langle I,I,\alpha|I,I,\alpha'\rangle=\delta_{\alpha\alpha'}
\end{equation}

 Thus the  set of functions (3) is orthogonal.
 Then, it follows from the above calculations that the operators $I_{\pm}$,
 for this set of functions, have no nonzero matrix elements between the
 functions with different $\alpha$.
 Vanishing of the matrix element of the type
$\langle I,L,\alpha|I_\pm|I,L\mp 1,\alpha'\rangle$ for $\alpha\ne\alpha'$
 can be proved by the procedure similar to Eqs. (4) -- (8).
  At $\alpha=\alpha'$ the values of these matrix elements are known from the
 theory of angular momentum \cite{Lan}:

\begin{equation}
\langle I,L,\alpha|I_+|I,L- 1,\alpha\rangle=
\langle I,L-1,\alpha|I_-|I,L,\alpha\rangle=
\sqrt{I(I+1)-L(L-1)}
\end{equation}

Let us examine the matrix of Hamiltonian (1) in the representation of the
functions obtained by multiplying the functions (3) (corresponding to the nuclear degrees of freedom)
 by the wave functions  of electron spin with a given $z$-projection
 $|S\rangle, S=\pm 1/2$: $|S,I,L,\alpha\rangle\equiv|S\rangle|
I,L,\alpha\rangle$.
 First, as it follows from the above results, this matrix can
be divided into blocks with different total nuclear angular momenta $I$.
 The dimension of each of the blocks is given by:
(the number of possible projections of the total nuclear angular
momentum)$\times$ (the number of possible projections of the electron spin)$=(2I+1)2$.
 The number of these blocks with a given total angular momentum of the nuclei is
equal to the number of ways how the angular momentum $I$ can be obtained by
summarizing $2N$ elementary nuclear angular momenta 1/2.
We denote this number  by $\Gamma_N(I)$.
 It can be shown (the idea how one can do this is presented in\cite{Sch}) that

\begin{equation}
\Gamma_N(I)=C_{2N}^{N-I}-C_{2N}^{N-I-1}\hskip10mm I=0,1,...,N.
\end{equation}
 Thus to diagonalize Hamiltonian (1) it suffices to diagonalize each block with a
given nuclear angular momentum $I$.
 Let us order the basis functions of a chosen block in the following way:
$|+,I\rangle,|-,I\rangle,|+,I-1\rangle,
...|-,1-I\rangle,|+,-I\rangle,|-,-I\rangle$.
 One can easily see that, for the basis functions arranged in this way, the
considered block of the Hamiltonian appears to be, in turn, divided into
blocks, with two of them, corresponding to the states $|\pm,\pm I\rangle$,
being one dimensional and 2$I$ of them, two dimesional.
 General form of the two dimesional block corresponding to the states
  $|-,L\rangle$ $|+,L-1\rangle$, where $L=I,I-1,...,1-I$, is

\begin{equation}
H_{LI}=\left(\matrix{-(\omega+A_\| L)/2 & A_\bot\sqrt{I(I+1)-L(L-1)}\cr
A_\bot\sqrt{I(I+1)-L(L-1)} &  (\omega+A_\| (L-1))/2}\right)
\hskip5mm L=I,I-1,...,1-I
\end{equation}

 Diagonalization  of the two dimensional blocks can be easily performed
  and, as a result, for the final  form of the spectrum of Hamiltonian (1) we
  have:

\begin{equation}
E_{LI}^\pm={1\over 2}\bigg(
-{A_\|\over 2}\pm\sqrt{
[\omega+A_\|L-A_\|/2]^2+4A_\bot^2[I(I+1)-L(L-1)]
}\bigg),\hskip3mm L=I,...,1-I
\end{equation}
$$
E_I^\pm={A_\|I\pm\omega\over 2},\hskip10mm I=0,1,...,N
$$

The multiplicity of degeneracy of each eigen-energy is given by Eq. (10).

\section{Dynamics of the electron spin}

As was mentioned in introduction, Hamiltonian (1) can describe spin dynamics of
a quantum dot.
 In the simplest way, the typical experiment can be represented as follows.
 A short polarized laser pulse transverse the quantum dot into the excited
 state with the spin-oriented excited electron.
 Then, the spin dynamics of this electron is detected by time dependence of
 polarization of its luminescence.
  This dynamics is considered to be determined by the contact interaction of the
 electron spin with the nuclei of the quantum dot.
 Hamiltonian (1) corresponds to the assumption that the excited electron wave
function uniformly covers all the nuclei of the quantum dot.
 Bearing in mind this experimental arrangement let us assume that at the time
moments  $t=T_m, m= 0,1,2,3,...$ the system is subjected to the action of short
high-power optical pulses which orient the electron spin. Now the problem can
be mathematically formulated in the following way:

1. Right after the pulse, the electronic density matrix is assumed to have the
form:
$$
\rho^e(T_m+0)=\hbox{Sp}_I\hskip1mm \rho(T_m+0)=\left(\matrix{\cos^2\phi& 0\cr\ 0&
\sin^2\phi}\right)
$$
 where the symbol  Sp$_I$ denotes diagonal summation of the total density matrix
 before the pulse  $\rho(T_m+0)$ over the nuclear quantum numbers.
 The angle $\phi$ quantifies the degree of orientation of the electron spin by
the optical pulse and depends on the pulse polarization.
 Thus the observables associated with the electron spin at $t=T_m$, generally,
  experience a jump.

2. We assume that all the observables associated with the nuclear system are, on
the contrary, continuous  at $t=T_m$, i.e., that they are not directly affected
by the optical pulse.
 It means that the nuclear density matrix
\begin{equation}
\rho^n=\hbox{Sp}_S\rho
\end{equation}
 is continuous at $t=T_m$, i.e.,
  $\rho^n(T_m-0)=\rho^n(T_m+0)\equiv\rho^n(T_m)$.
  The symbol Sp$_S$ in (13) denotes diagonal summation of the total density
 matrix over electronic quantum numbers.

3. The above requirements imposed on the electronic and nuclear density matrices
are, generally, not enough to unambiguously reconstruct the total density
matrix of the system after the pulse.
 To do that, one has to consider the dynamics of the system during the pulse.
 This, however, lies outside the scope of this paper.
 Below, we make the simplest assumption compatible with the above requirements
 that the total density matrix immediately after the pulse can be represented
 as a product of the nuclear density matrix before the pulse and the electronic
 density matrix corresponding to the oriented electron spin:

\begin{equation}
\rho(T_m+0)=\rho^n(T_m)\left(\matrix{\cos^2\phi& 0\cr\ 0&
\sin^2\phi}\right)
\end{equation}

4. We assume that at the very beginning of the experiment the nuclear density
matrix corresponds to equi-probable distribution over all  nuclear
states (high-temperature density matrix).

5. We seek for the dynamics of the system under the action of a train of optical
 pulses.

The problem formulated in this way can be consistently solved.
 First of all we will show that the nuclear density matrix, under the
assumptions made above, is always diagonal in the representation of functions with
 specified total nuclear angular momentum and its projection.
 Let at $t=0$, the system to be excited by the first optical pulse that orients
 the electron spin.
 Then, at $t=+0$ the electron spin appears to be  $1/2(\cos^2\phi-\sin^2\phi)$,
while the nuclear state remains unperturbed (high-temperature state).
 Let us  write the appropriate density matrix (14) in the representation
of the functions $|S,I,L,\alpha\rangle$ used in the previous section:

\begin{equation}
 \langle S,I,L,\alpha|\rho|S',I',L',\alpha'\rangle=
2^{-2N}\delta_{SS'}\delta_{II'}\delta_{LL'}\delta_{\alpha\alpha'}
\cases{\cos^2\phi \hskip3mm\hbox{при}\hskip3mm S=1/2\cr \sin^2\phi\hskip3mm \hbox{при}
\hskip3mm S=-1/2}
\end{equation}
 In accordance with the above results Hamiltonian (1) can be divided into blocks
 of total nuclear angular momenta $I$ (referred to as $I$-blocks), with two one
 dimensional and $2I$ two-dimensional blocks within each of them.
 We will call last-mentioned two dimensional blocks $LI$-blocks, with $L$ being
  the projection of the total nuclear angular momentum corresponding to the
  upper-left corner of the block (11).
 To have a  clearer idea about the topography of the  matrices considered in the
problem under study, we show in fig.1 the block corresponding to the states with
$I=1$.
 The symbols $"+"$ and $"-"$ denote nonzero elements of the density matrix (15)
in this block, related to the electron spin projections $+1/2$ and $-1/2$,
respectively.
The nonzero elements of Hamiltonian (1) are colored gray.
 Since the initial density matrix (15) is diagonal, the nonzero elements of the
density matrix may appear, in process of its evolution only where the matrix
elements of the Hamiltonian are nonzero, i.e, in the places colored gray.
  Thus the total density matrix under the initial condition (15), will have the
 topography identical to that of the Hamiltonian, i.e, it will be
 block-diagonal.
 Now let us assume that the system under consideration is acted upon by the
 second optical pulse polarizing the electron spin.
 According to the formulation of the problem described above the density matrix
 of the system will have the form Eq.(14).
 To obtain the element $\rho^n_{LL'}$ of the nuclear density matrix, entering
Eq.(14), we have to sum up the elements of the total density matrix on the
diagonal of the two dimensional block corresponding to these nuclear states.
 For instance to obtain the matrix element $\rho^n_{1,-1}$ (fig.1) we have to
perform diagonal summation in the upper-right block (the relevant matrix elements of the total density
matrix are located in the crossed cells)
 All the blocks whose diagonal summation is needed to obtain the nuclear
 density matrix are separated in fig.1 by thick lines.
 Since the nonzero matrix elements of the total density matrix are located in
the cells colored gray, one can see from the fig.1 that the nuclear density
matrix after the second pulse will be diagonal in the representation of the
total nuclear angular momentum and its projection.
 Using similar reasoning to the third, forth etc. pulses, we obtain (by
induction), the nuclear density matrix, for the problem under consideration, is
always diagonal in the representation of the
total nuclear angular momentum and its projection.
  Since the hole problem brakes down in to blocks of the total nuclear angular momentum
 $I$ we can denote the diagonal elements of the nuclear density matrix of
 the block $I$ by
   $U_{LI}, L=-I,1-I,...,I-1,I$.
 To completely specify the block, one should have indicate its quantum number
 $\alpha$ of the used representation. However, we shall not present it because
 no observable depends on this number.
 Inside the block with a given $I$, one can consider the two-dimensional $LI$-blocks
 corresponding to those of Hamiltonian $H_{LI}$ (11).
 The corresponding $LI$-blocks of the density matrix, which are further denoted
by $\Theta_{LI}$, meet the equation

   \begin{equation}
   \imath \dot{\Theta}_{LI}=[H_{LI},\Theta_{LI}]
   \end{equation}

  The quantum-mechanical problems related to two dimensional matrices are
 well known and, in what follows, we use the appropriate terminology.
  It is convenient to express the matrix $H_{LI}$ in terms of the Pauli matrices
 $\sigma_{x,y,z}$:

  \begin{equation}
  H_{LI}=-{A_\|\over 4}+b_{LI}\sigma_z + c_{LI}\sigma_x
  \end{equation}
where
  \begin{equation}
  b_{LI}=A_\|(1/2-L)-\omega \hskip20mm   c_{LI}=2A_\bot\sqrt{I(I+1)-L(L-1)}
  \end{equation}

The sought density matrix $\Theta_{LI}$ can be also expressed in terms of the
unit matrix (with the coefficient denoted by  $\theta^{LI}_0$ ) and the Pauli
matrices

  \begin{equation}
\Theta_{LI}=\theta^{LI}_0+\theta^{LI}_x\sigma_x+\theta^{LI}_y\sigma_y+
\theta^{LI}_z\sigma_z
  \end{equation}

Here, the quantities  $\theta^{LI}_{x,y,z}$ are the components of so the
so-called Bloch vector  \cite{Abr}.

Then, the standard solution of the problem of dynamics of a two-level system
 yields for the Bloch vector the following expression:

\begin{equation}
{\bf\theta}^{LI}(t)={\bf M}^{LI}(t) \theta^{LI}(0)
\end{equation}

Where the matrix ${\bf M}^{LI}$ is defined as:
\begin{equation}
{\bf M}^{LI}(t)=\left(\matrix{1+{b^2\over\Omega^2}(\cos\Omega t-1)&-{b\over\Omega}\sin\Omega t&{b c\over\Omega^2}(1-\cos\Omega t)\cr
{b\over\Omega}\sin\Omega t&\cos\Omega t&-{c\over\Omega}\sin\Omega t\cr
{b c\over\Omega^2}(1-\cos\Omega t)&{c\over\Omega}\sin\Omega t&1+{c^2\over\Omega^2}(\cos\Omega t-1)}\right)
\end{equation}

$$
b=b_{LI}, \hskip10mm c=c_{LI},\hskip10mm \Omega=\Omega_{LI}=\sqrt{b^2+c^2}
$$

Let the optical pulse create the electron in a partially polarized state with
the density matrix

\begin{equation}
\left(\matrix{\cos^2\phi &0\cr 0& \sin^2\phi}\right)
\end{equation}

At $\phi=0 \hskip2mm(\pi/2)$, the value of the  electron spin is $1/2 \hskip2mm(-1/2)$.

For the initial state of the type (14), where the nuclear density matrix has
only diagonal elements $U_{LI}$, the initial condition for the block $LI$ of
the total density matrix has the form:
\begin{equation}
\Theta_{LI}(0)=
\left(\matrix{ U_{LI}\sin^2\phi &0\cr 0&  U_{L-1,I}\cos^2\phi}\right)
\end{equation}

ƒл€ начального вектора Ѕлоха и величины $\theta_0$ блока $LI$ это дает:
For the initial values of the Bloch vector and the quantity $\theta_0$ related to the block
$LI$, it yields:
\begin{equation}
\theta^{LI}(0)=\left(\matrix{0\cr 0\cr  U_{LI}\sin^2\phi-
U_{L-1,I}\cos^2\phi}\right),\hskip5mm \theta_0^{LI}={1\over 2}
(U_{LI}\sin^2\phi+U_{L-1,I}\cos^2\phi)
\end{equation}

Contribution of the block $LI$ to the mean value of the
electron spin is $-\hbox{Sp}\hskip1mm\Theta_{LI}(t)\sigma_z$.

Its convenient to use the double value of this contribution, which is denoted
as

\begin{equation}
\Delta_{LI}(t)=-2\hbox{Sp}\hskip1mm\Theta_{LI}(t)\sigma_z=
-\theta^{LI}_z
\end{equation}
It follows from Eqs.(20),(21) and (24) that
\begin{equation}
\Delta_{LI}(t)=\bigg(U_{L-1,I}\cos^2\phi -U_{LI}\sin^2\phi \bigg)
\bigg(
1+{c^2_{LI}\over\Omega_{LI}^2}(\cos\Omega_{LI} t-1)
\bigg)
\end{equation}

When calculating contribution of the block with a given total angular momentum
$I$, we have to take into account the two one-dimensional blocks, which provide
the following time-independent contribution to the mean value of the electron
spin: $( U_{I,I}\cos^2\phi- U_{-I,I}\sin^2\phi)/2$.
 Then the general formula for the time dynamics of $z$-projection of the electron
spin prepared , at $t=0$, in the state (22) for the initial density matrix $U_{LI}$
 is given by:
\begin{equation}
\langle S(t)\rangle={1\over 2}\sum_{I=0}^N \Gamma_N(I)\bigg(
\sum_{L=1-I}^I\Delta_{LI}(t)+U_{I,I}\cos^2\phi- U_{-I,I}\sin^2\phi
\bigg)
\end{equation}
где:
$$
\Delta_{LI}(t)=\bigg(U_{L-1,I}\cos^2\phi -U_{LI}\sin^2\phi \bigg)
\bigg(
1+{c^2_{LI}\over\Omega_{LI}^2}(\cos\Omega_{LI} t-1)
\bigg)
$$
$$
 b_{LI}=A_\|(1/2-L)-\omega, \hskip7mm   c_{LI}=2A_\bot\sqrt{I(I+1)-L(L-1)},
 \hskip7mm\Omega^2_{LI}=b_{LI}^2+c^2_{LI}
$$
$$
\Gamma_N(I)=C_{2N}^{N-I}-C_{2N}^{N-I-1}
$$

It is often interesting to know the steady state value of the electron spin
polarization.
 This value can be obtained by removing the time-dependent component in each of the
contributions (27), i.e., using the substitution
   \begin{equation}
   \Delta_{LI}\rightarrow\Delta_{LI}^{st}=
   \bigg(U_{L-1,I}\cos^2\phi -U_{LI}\sin^2\phi \bigg)
\bigg(
1-{c^2_{LI}\over\Omega_{LI}^2}
\bigg)=
\bigg(U_{L-1,I}\cos^2\phi -U_{LI}\sin^2\phi \bigg)
{b^2_{LI}\over\Omega_{LI}^2}
   \end{equation}

For the case  of (i) the  isotropic contact interaction $2A_\bot=A_\|\equiv A$,
 (ii) zero magnetic field $\omega=0$, (iii)  circularly polarized optical pulse
 $\phi=0$, (iv) high-temperature initial nuclear density matrix $U_{LI}=2^{-2N}$,
 and (v) $2N\gg 1$, for the electron spin dynamics, one can derive a  simple
 formula.

To do that, note that:

1.the frequencies $\Omega_{LI}=AI$, in this case, do not depend on the quantum
number of projection of the total angular momentum,

2. at $2N\gg 1$ we may assume that $c_{LI}^2=A^2(I^2-L^2)$,

3. for  $\Gamma_N(I)$ given by  (10), we can use the asymptotic expression

\begin{equation}
\Gamma_N(I)=-{2^{2N}\over\sqrt{\pi N}}\hskip1mm{d\over dI}
\exp\bigg(-{I^2\over N}\bigg)
\end{equation}

Then, by substituting in Eq.(27) the summation by integration and calculating
the integrals, we obtain

\begin{equation}
\langle S(t)\rangle={1\over 6}+{1\over 3}\bigg(
1-{NA^2t^2\over 2}\bigg)
\exp\bigg(-{NA^2t^2\over 4}\bigg)
\end{equation}

 A formula similar to (30) was derived based on intuitive
 considerations about the nuclear field and its statistics
  in \cite{Mer}.
 As seen from Eq.(27), the electron spin dynamics is controlled by the nuclear
state of the system, which, in turn, can vary under the repetitive orientation
of the electron spin by  optical pulses.
 So in the next Section we will derive the laws of
 pulse-to-pulse transformation of the nuclear density matrix
  and will calculate the steady-state nuclear density matrices
   for different regimes of the repetitive optical orientation of the electron
   spin.

 \section{Nuclear dynamics}

\subsection{Transformation of the nuclear density matrix under optical
orientation}

As was shown in the previous section, the density matrix blocks corresponding
to states with a given total angular momentum $I$ are mutually independent.
So, in what follows we will consider only one of the blocks, for which
 we formulate the following problem:

1.
Let our system be repetitively (with the period $T$) acted upon by an optical pulse
 which  orients the electron spin  creating the states of the type (14) at the
 moments $T_m=mT$.

2. Let the block of the nuclear density matrix corresponding to the states with
the total angular momentum $I$ have to the moment of arrival of the $n$-th pulse, the
diagonal elements $U_{LI}^n$.

3. Find the elements of this block  $U_{LI}^{n+1}$ to the moment of arrival of the $(n+1)$-th
pulse.

 As was already mentioned, the nuclear density matrix, in our case, is always
 diagonal, and to obtain its elements, one has to sum up diagonal elements of
 the diagonal two-dimensional block with a given value of the total nuclear
 angular momentum projection and with different projections of the electron
 spin.
 Note that these blocks do not coincide with the blocks $LI$.
 In fig.1, these blocks are separated by thick lines, while the blocks $LI$ are
 colored gray.
 Keeping this in mind, we can express the diagonal elements of the nuclear
density matrix in terms of the elements of $LI$-blocks of the density matrix
and the parameters defined by (19):

\begin{equation}
U_{LI}=\Theta_{22}^{L+1,I}+\Theta_{11}^{LI}=
\theta_0^{L+1,I}+\theta_0^{LI}-{1\over 2}(\theta_z^{L+1,I}-\theta_z^{LI})
\end{equation}

For the extreme values $L=\pm I$, we have to assume that

\begin{equation}
\Theta^{I+1,I}_{22}\equiv U_{II}\hskip5mm
\Theta^{-I,I}_{11}\equiv U_{-II}
\end{equation}

 Using the results of the previous section and introducing the quantities

\begin{equation}
\Phi_{LI}\equiv {1\over 2}\bigg({c_{LI}\over \Omega_{LI}}\bigg)^2
\bigg[
\cos(\Omega_{LI}T) - 1\bigg],
\end{equation}

 we obtain the following formulas for the pulse-to-pulse transformation
  of the nuclear density matrix:

\begin{equation}
U_{LI}^{n+1}=
U_{LI}^n[1+\Phi_{LI}\sin^2\phi+\Phi_{L+1,I}\cos^2\phi]-
\end{equation}
$$
-U_{L-1,I}^n\Phi_{LI}\cos^2\phi-
U_{L+1,I}^n\Phi_{L+1,I}\sin^2\phi
+{U_0-U_{LI}^n\over\tau}
\hskip10mm \hbox{при} \hskip3mm L=I-1,...,1-I
$$
For the the extreme values $L=\pm I$:

\begin{equation}
U_{II}^{n+1}=U_{II}^n+\Phi_{II}(U_{II}^n\sin^2\phi-U_{I-1,I}^n\cos^2\phi)
+{U_0-U_{II}^n\over\tau}
\end{equation}
\begin{equation}
 U_{-II}^{n+1}=U_{-II}^n-\Phi_{1-I,I}( U_{1-I,I}^n\sin^2\phi- U_{-II}^n\cos^2\phi)
+{U_0-U_{-II}^n\over\tau}
\end{equation}

 The last terms in Eqs. (34) --(36) take into account the possible nuclear
 relaxation, with $\tau$ being the relaxation time in the units of $T$ and
 $U_0=2^{-2N}$ being the diagonal elements of the high-temperature nuclear
 density matrix.

\subsection{The steady-state  nuclear density matrix for the case of repetitive
 sign-constant orientation of the electron spin.}

 If the angle $\phi$ does not change from pulse to pulse, then after
 sufficiently long time, the nuclear density matrix will stabilize
   $\lim_{n\rightarrow\infty}U_{LI}^n=U^{st}_{LI}$.
 The equations for the steady-state nuclear density matrix can be obtain from
the conditions $U_{LI}^{n+1}=U_{LI}^n=U_{LI}^{st}$

\begin{equation}
U_{LI}^{st}[\Phi_{LI}\sin^2\phi+\Phi_{L+1,I}\cos^2\phi]
-U_{L-1,I}^{st}\Phi_{LI}\cos^2\phi-
U_{L+1,I}^{st}\Phi_{L+1,I}\sin^2\phi
+{U_0-U_{LI}^{st}\over\tau}=0
\end{equation}
$$
\hbox{при} \hskip3mm L=I-1,...,1-I
$$
 For the extreme values,  $L=\pm I$:
\begin{equation}
\Phi_{II}(U_{II}^{st}\sin^2\phi-U_{I-1,I}^{st}\cos^2\phi)
+{U_0-U_{II}^{st}\over\tau}=0
\end{equation}
\begin{equation}
-\Phi_{1-I,I}( U_{1-I,I}^{st}\sin^2\phi- U_{-II}^{st}\cos^2\phi)
+{U_0-U_{-II}^{st}\over\tau}=0
\end{equation}

Since the blocks $I$ are independent the condition of normalization

 \begin{equation}
 \sum_{L=-I}^IU_{LI}=(2I+1) U_0\hskip20mm I=0,1,2,...,N
 \end{equation}
  should be met.
Normalization of the total nuclear density matrix has the form:
\begin{equation}
\sum_{I=0}^N\Gamma_N(I)\sum_{L=-I}^I U_{LI}=1
\end{equation}

 It can be easily verified that in the absence of relaxation $\tau\rightarrow\infty$ and
и $\phi=0$ ($\phi=\pi/2$), the steady state nuclear density matrix corresponds
to the complete polarization of the nuclear angular momentum in the block $I$:
 $U_{LI}^{st}=0, L\ne I (-I)$ и
 $U_{II}^{st}(U_{-II}^{st})=U_0(2I+1)$.
(Note that this does not correspond to the total polarization of all the
nuclei. The latter state would have corresponded to the nuclear density matrix
with all blocks $I$ empty except for the one with $I=N$
 with the only populated state with the grates magnitude of the projection $L=\pm N$).
 When the pumping is performed with linearly polarized (or non-polarized)
 pulses (i.e., $\phi=\pi/4$) the steady-state density matrix, as can be shown
 by direct calculations, remains high-temperature $U_{LI}=U_0$.
 In the case $\tau\neq\infty$ and complete optical orientation (for example when $\phi=0$)
 the Eqs. (37) -- (39) allows one  to rapidly calculate, in
a recurrent way, the steady state  nuclear density matrix.
 For that one may use the following steps:

1. using (39) calculate $U^{st}_{-II}=1$.

2. using Eq. (37), successively calculate $U^{st}_{LI}$, $L=1-I,...,I-1$.

3.  using Eq.(38), derive $U^{st}_{II}$

Note that the above procedure of calculation of the steady-state nuclear density
matrix and 300 points of the magnetic-field dependence of the steady-state electron spin
polarization,
 for example for a system comprised of 1600 nuclei, for the case of
repetitive optical orientation of the electron spin takes, when using an up-to-date
computer, just a few seconds.
 The dynamics can be calculated even faster, because the steady-state density
matrix is the same for all the time dependence and may be calculated only once.
 In the numerical calculations, one should keep in mind the following.
 Examination of the function $\Gamma_N(I)$ (29) shows that its width and position
of its maximum $\sim\sqrt{N}$, i.e., $\Gamma_N(I)$ is essentially non-zero in
the region $I\sim \sqrt{N}$.
 For this reason it is useful, in the numerical calculations to find the
position of the maximum and, when calculating the sums (27), to restrict
oneself to region $I<(3 \div 4) I_{max}$.

 Equations (34) --(36) allow one to calculate the pulse-to-pulse dynamics of the density
 matrix (with all the observables dependent on the nuclear degrees of freedom).
 For instance, the nuclear angular momentum projection $\langle I_z\rangle$ can be calculated using
 the formula

\begin{equation}
\langle I_z\rangle=\sum_{I=0}^N\Gamma_n(I)\sum_{-I}^I U_{LI}L
\end{equation}

 Note here that the maximum nuclear polarization (for the considered action upon the system)
  is obtained for total polarization inside the blocks with a given $I$, i.e.,
$U_{LI}=0, L\ne I$  and  $U_{II}=U_0(2I+1)$.
 Therefore the maximum projection of the nuclear moment, in our case, is given
 by the formula
   \begin{equation}
\langle I_z\rangle_{max}=U_0\sum_{I=0}^N\Gamma_N(I)(2I+1)I\approx 2\sqrt{N\over\pi}
   \end{equation}

\subsection{Examples of calculating the spin dynamics and the steady-state spin
polarization for the sigh-constant repetitive spin orientation}

 Now, we will present some examples that show that the steady-state
polarization and the electron spin dynamics substantially depend on the way of
preparation of the nuclear system and considerably differ on those calculated
 using the high-temperature nuclear density matrix.
 Figure 2(a) shows magnetic-field dependences of the steady-state electron spin
polarization under repetitive complete optical orientation for different
nuclear relaxation times $\tau$ (curves 1 -- 4).
 The same figure shows the results of calculations for the case of high-temperature
nuclear density matrix (curve 0).
 As the nuclear relaxation time increases, the curves differ more and more from
  the case of the high-temperature nuclear density matrix:
  there arise an asymmetry related to the sign of the optical orientation
   and a dip in small magnetic fields.
 Figure 2(b) shows a family of magnetic field dependences of the nuclear
polarization  $\langle
 I_z\rangle/N$ corresponding to figure 2(a).
 The electron spin dynamics after a long repetitive orientation, in the
 presence of the nuclear relaxation, is shown in figure 3.
  This dynamics, as seen from the figure, is quasi-periodic with the period of
 the optical orientation $T$.
  The dynamics in figure 3 corresponds to the case when the train of the pulses,
 after its long action upon the system, is switched off and one can observe the
 spin dynamics infinitely long.
  If the orienting pulses are not interrupted, then, in the described behaviour
  the spin polarization goes ahead of the orienting pulse, which in some cases,
  can be indeed observed experimentally.

\subsection{The case of a  repetitive sign-alternating spin orientation}

 Formulas (34) --(36) are the linear transformation of the vector-column
 ${\bf U}_I$ of the nuclear density matrix from $n$-th to $n+1$-th pump pulse.
 In the general case, polarization of the orienting pulse may change from pulse
 to pulse, i.e. $\phi=\phi_n$.
 Then, the above linear transformation can be represented in the matrix form:
 \begin{equation}
 {\bf U}_I^{n+1}={\bf T}_I(\phi_n) {\bf U}_I^n
 \end{equation}
 The transformation matrix ${\bf T}_I$ is given by formulas (34) -- (36).
  When the nuclear relaxation is weak, then in the above case of the complete repetitive  sign
 constant orientation, one should expect a strong nuclear polarization (inside the blocks with a given
 total nuclear angular momentum) with decreasing signal of the electron spin
 dynamics because the electron spin, in this case, is conserved.
  This is the reason why the sign alternating  orientation of the electron spin
 is often used in the experiments.
 In simplest case it means that e.g., for the even pulse, $\phi=0$ and for the
odd pulse $\phi=\pi/2$.
 In this case the nuclear density matrix will always change from pulse to pulse,
 but, after sufficiently long time, the matrix arising after each even pulse (or after each odd pulse)
  will remain the same.
 This matrix evidently satisfies the condition
\begin{equation}
{\bf U}_I={\bf T}_I(0){\bf T}_I(\pi/2) {\bf U}_I
\end{equation}
  and can be calculated in the same way as it was made above for the case of the
 sign constant orientation.
 When the nuclear relaxation is absent, the formulas for the non-normalized
  steady-state (in the sense of Eq.(45)) nuclear density matrix have the form:

\begin{equation}
U_{1-I}=U_{-I}{1+\Phi_{1-I}\over 1+\Phi_{2-I}}
\end{equation}
\begin{equation}
U_{L+1}=
{U_L[(1+\Phi_L)(1+\Phi_{L+1})+\Phi_{L+1}^2-1]-U_{L-1}(1+\Phi_L)\Phi_L
\over (1+\Phi_{L+2})\Phi_{L+1}},
\hskip5mm L=1-I,...,I-2
\end{equation}
\begin{equation}
U_I=U_{I-1}(1+\Phi_I)
\end{equation}
 The recurrent procedure and normalization are completely similar to those
 described above.
 For this regime of the orientation, the spin dynamics also exhibits a
quasi-periodic behaviour (fig. 4a), whereas the magnetic field dependence of
the steady-state electron spin polarization is getting symmetrical but may show
a dip in region of low magnetic fields. (fig.4b)
The  curve in fig.4b with no dip corresponds to the case of  the
high-temperature nuclear density matrix.

\section{The electron-nuclear spin echo}

As seen from the aforesaid, the system under study is equivalent to an ensemble
 of two-level systems.
  Each two-dimensional $LI$-block of the Hamiltonian may be put into
 correspondence with a two-level system with the energy gap $\Omega_{LI}$.
 The  contribution of each $LI$-block to the electron spin $z$-projection is
additive and is determined by formula (27).
  On the other hand, this contribution is related to the $z$-component of the
 Bloch vector $\theta^{LI}_z(t)$ (19).
 For these reasons the system under study should show all the effects inherent
of an ensemble of two-level systems in the presence of inhomogeneous broadening.

 For example, the effect of arising replicas in spin dynamics (fig.3), mentioned
 above, has much in common with the effects described in \cite{Hes,Mor,Che}.
 Consider now the effect of spin echo \cite{Abr} in such a system.
 Below, we present a standard description of this effect keeping in mind a
 possible experiment with a quantum dot.
  Let the electron spin, at $t=0$, is optically polarized along $z$-direction in
 zero magnetic field.
 Consider the motion of the Bloch vector corresponding to some $LI$-block.
 Assume, for simplicity, that  $A_\|=0$.
 Then the motion of the Bloch vector ${\bf \theta}_{LI}$ of the $LI$-block (11)
  represents rotation around the $x$-axis with the frequency $\Omega_{LI}$.

 The Bloch vectors pertaining to different $LI$-blocks are rotated with
 different frequencies, so that the vectors  ${\bf \theta}_{LI}$ initially
 oriented along $z$-axis, after sufficiently long time span, will spread more
 or less uniformly over the plane perpendicular to the $x$-axis.
 \footnote{ In the considered case of $A_\|=0$, it corresponds to vanishing of the
  electron spin $z$-projection.
 At $A_\|\ne 0$, it is not the case, and the dynamics of the Bloch vectors $\theta^{LI}$
 appears   to be more complicated, but it is not important for the qualitative
 interpretation of the echo effect}
 After it happened, let us apply to the system a $\pi$-pulse of the magnetic
 field directed along the $z$-axis.
 In this pulse the magnitude of the magnetic field $\omega$ and its duration $\Delta
   T$ should meet the condition $\omega\Delta T=\pi$.
 Denote the initial moment of the $\pi$-pulse by $T_1$.
  We assume, for simplicity, that the magnetic field is so strong that
  $\omega\gg 2A_\bot\sqrt{I(I+1)-L(L-1)}$ and that the dynamics of the Bloch
  vectors of the $LI$-blocks, during the pulse, is predominantly  determined by
  the magnetic field and, therefore, represents rotation around the $z$-axis.
 Thus during interval, $\Delta T$ all the Bloch vectors will rotate around the
$z$-axis by the angle $\pi$ and the vector with the greatest frequency $\Omega_{LI}$
which is, at the beginning of the $\pi$-pulse, ahead of all other vectors,
after the pulse will be behind all other vectors.
 On the contrary, the vector with the minimum frequency $\Omega_{LI}$
 which is, before the pulse, behind all other vectors, after the pulse will be ahead
 of them.
 After the end of the $\pi$-pulse the fastest Bloch vector, which is now behind
all others, will start to reach the slowest vector which is now ahead of others
 and after the time interval  $T_1$ after the pulse they will meet, both being
 oriented, at this moment, along $z$-axis, as at $t=0$.
 The Bloch vectors between the fastest and the slowest ones will also come
 together after the time interval $T_1$ after the end of the pulse.
 Thus at $t=2T_1+\Delta T$, the electron spin projection will be equal to that
 at $t=0$
 This is what is referred to as the echo signal.
 The simplifications adopted above do not qualitatively change the picture of
 the effect.
 It is also clear that instead of the $\pi$-pulse one can use a pulse
 corresponding to odd number of half-rounds of the electron spin.
 Formally, the echo signal is calculated using formulas (25)  and (27),
  with the needed Bloch vectors of the $LI$ blocks calculated using the
  formulas:
   \begin{equation}
   {\bf \theta}^{LI}(t)=\cases{{\bf M}^{LI}(0,t)\hskip1mm\theta^{LI}(0)\hskip65mm\hbox{при}\hskip3mm t<T_1\cr
   {\bf M}^{LI}(\omega,t-T_1)\hskip1mm{\bf M}^{LI}(0,T_1)\hskip1mm\theta^{LI}(0)\hskip36mm\hbox{при}\hskip3mm T_1+\Delta T>t>T_1\cr
{\bf M}^{LI}(0,t-T_1-\Delta T)\hskip1mm{\bf M}^{LI}(\omega,\Delta T)\hskip1mm{\bf M}^{LI}(0,T_1)
 \hskip1mm\theta^{LI}(0)
 \hskip3mm\hbox{при}\hskip3mm t>T_1+\Delta T
}
   \end{equation}
 Here the initial Bloch vector of the $LI$-block $\theta^{LI}(0)$ is given
 by Eq.(24) and, thus, depends on the nuclear density matrix formed to the
 moment of observation of the effect.
  The first and the second arguments  of the matrices determine there dependence
 on the magnetic field and time, respectively.
  Figure 5 shows the calculated effect of the $\pi$-pulse on the electron spin dynamics
 at different values of the pulse amplitude.
 One can see that the quality of the echo signal increases with increasing
 field amplitude because the dynamics of the system approaches the one
 described above.
 As is seen from the above treatment, the observation of the echo effect, in
 this case, does not require any inhomogeneous broadening and is possible for a
 single quantum dot.
The main difference between the technic described above and the standard method
 of spin echo is that an optical pulse is used instead of the $\pi/2$-pulse.

\newpage

\section*{Captures}

Fig.1 Topography of the matrix elements of the block with the total angular
momentum $I=1$. Non-zero matrix elements of the Hamiltonian are colored gray;
the blocks with given projection of the total nuclear momentum are separated by
thick lines.

Fig.2 (a) Magnetic-field dependences of the steady-state electron spin polarization
 under repetitive complete optical orientation ($\phi=0$) for different nuclear
 relaxation times (curves 1 --4). Curve (0) is calculated for the case of
 high-temperature nuclear density matrix.
(b) -- the corresponding family of  field dependences of the nuclear
polarization $\langle
 I_z\rangle/N$.

Fig.3 The electron spin dynamics after a long repetitive orientation of the elctron
 spin in the presence of nuclear relaxation.
  The arising replicas with the period equal to the orienting pulse repetition
  period is related to the arising odering in the nuclear density matrix.

Fig.4
Spin dynamics (a) and field dependence of the steady-state spin polarization
(b) for the case of sign alternating optical orientation and in the absence of
the nuclear relaxation.

 Fig.5 The echo signal in the electron spin dynamics for different values of the
$\pi$-pulse amplitude.
 The step-wise plots show time dependence of the magnetic field.

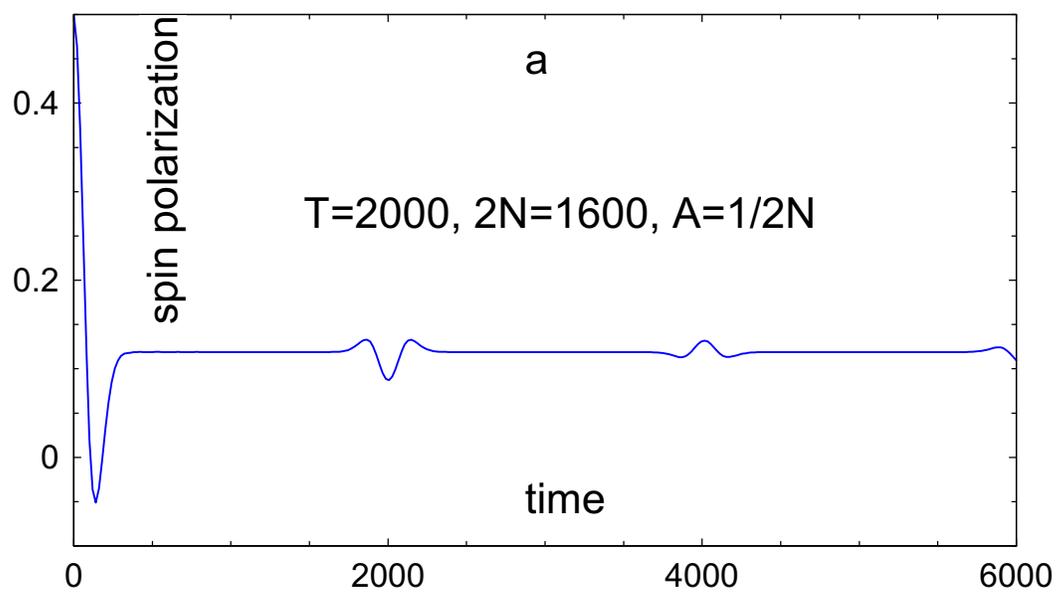
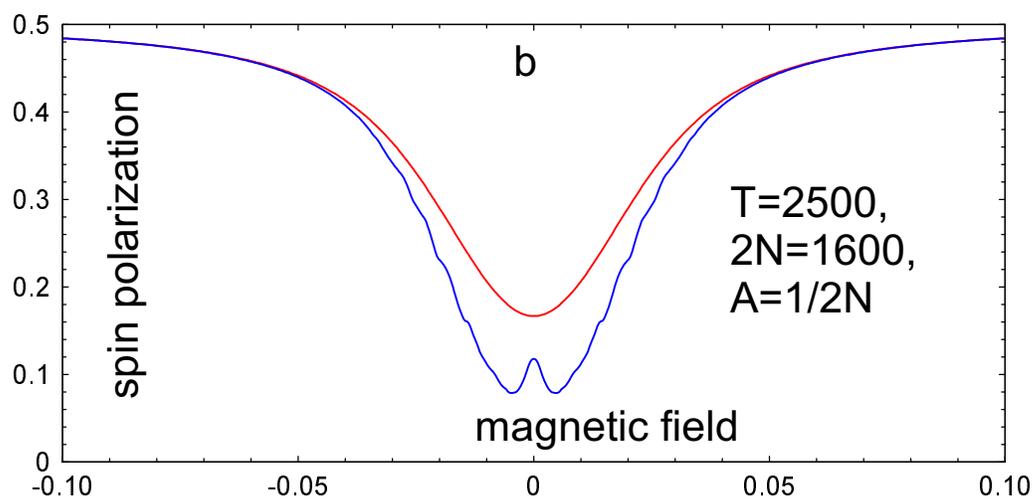
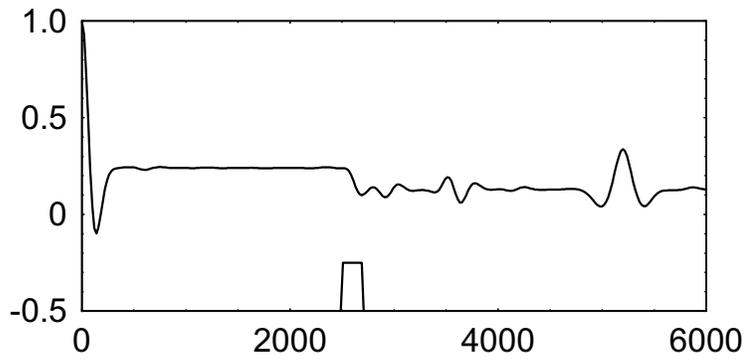
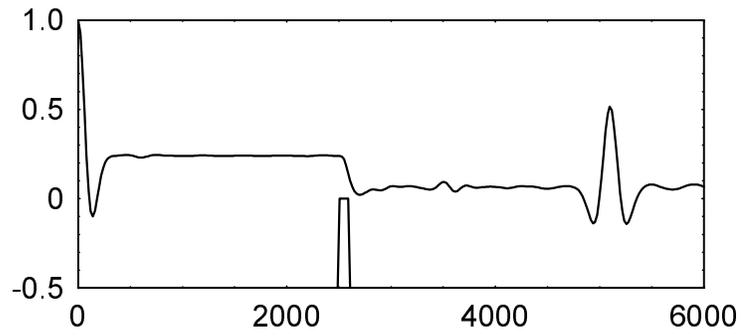
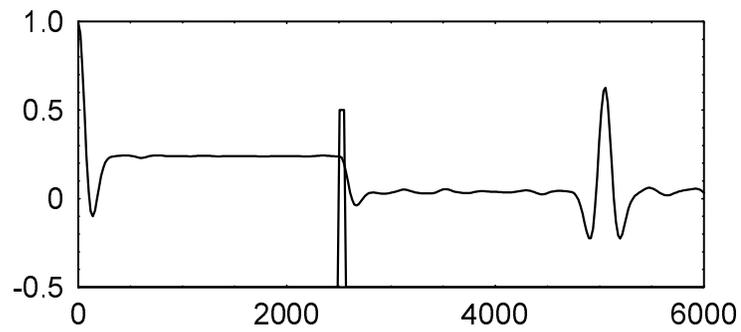
\end{document}